\documentstyle[12pt]{article}
\begin{document}
\begin{center}
\vspace{1.5in}
{\LARGE Neutron halo, tritons and alpha clusters in the 
nucleus ${^8}He$ }
\end{center}
\vspace{.4in}
\begin{center}
{\bf Afsar Abbas}\\
\vspace{.1in}
Institute of Physics\\ 
Bhubaneshwar-751005, India\\
email: afsar@iopb.res.in
\end{center}
\vspace{1.2in}
\begin{center}
{\bf Abstract}
\end{center}
\vspace{.3in}

Recent experiments clearly demonstrate the importance of the excited 
state $2^+$ at 1.8 MeV in the substruture  ${^6}He$ for the ground state 
of the nucleus ${^8}He$. It is shown here that this means that the ground 
state structure of ${^8}He$ rather than being $\alpha$ + 4n is actually
${^3}H$ + ${^3}H$ +2n. As such the pair of tritons would form a compact 
core around which the two valence neutrons provide a halo structure in 
${^8}He$. Hence both ${^6}He$ and ${^8}He$ are two neutron halo nuclei 
but with cores of quite different nature. All this demonstrates 
significance of triton clustering in neutron rich nuclei.

\newpage

The nuclei ${^6}He$ and ${^8}He$ are uniquely basic for the study of 
light neutron rich nuclei. While the nucleus  ${^6}He$ 
seems to have a structure
$\alpha$ + 2n, with the two neutrons making up a clear cut halo structure
around an inert $\alpha$ [1], the situation for 
${^8}He$ is known to be more complicated [1]. Though there does seem to be
a halo structure in  ${^8}He$, however one is not clear as to the nature 
of the same. 
Hence it is important to know as to what exactly is going on in ${^8}He$.

In focus here would be two very recent experiments
which give interesting insights about the structure of ${^8}He$.
These two very different and independent experimenst [2,3]
reveal identity of structure for ${^8}He$.
These seem to be ruling out the standard and canonical ( but somewaht
simplistic ) $\alpha$ + 4n structure of ${^8}He$.
Hence as such these two experiments cannot be understood in terms
of the conventional model of this nucleus. Here we shall discuss how a 
recent model proposed by the author
[4] is able to provide a consistent and revealing understanding of
the puzzles provided by these two experiments [2,3].

The first experiment [2] is that of the study of scattering of   
${^8}He$ on ${^4}He$ with the aim of detecting the 4n structure
of ${^8}He$. There have been some recent claims of observation of 
tetra-neutrons in light nuclei. This particular study was done with 
the aim of understanding this putative tetra-neutronic stucture in 
${^8}He$. They start with the standard assumption that $\alpha$ 
-particles form a well defined and inert object in ${^8}He$.
They considered a one step direct 4n transfer process and a two step 
sequential transfer process of 2n clusters proceeding through
the intermediate ground state and the first excited $2^+$ state
( at 1.8 MeV) in ${^6}He$. Their DWBA calculations indicated that the two 
step 2n transfer is clearly more important than the one step 4n transfer. 
This expriment therefore cleary disfavours the $\alpha$ + 4n cluster 
structure of ${^8}He$ [2].

In the second expermiment [3] they studied the two neutron transfer 
reactions $ p ({^8}He, t) {^6}He{_{gs}}$
and $ p ({^8}He, t) {^6}He{^*}({2^+})$
for the ground state and the first excited states of ${^6}He$  
respectively. One expects that that if the ground state of 
 ${^8}He$ contains the subsystem ${^6}He$ mainly in the $0^+$
state, then this would be preferencially populated in  
 ${^8}He (p,t) $ reaction. Then it being a second order proces, the 
cross section for the population of
 ${^6}He{^*} ( {2+} )$ would be considearbly lower. Contrary to these 
expectations they found the cross section of the
 $ p ({^8}He, d) {^6}He{^*}({2^+})$
reaction to be much higher than that of 
$ p ({^8}He, d) {^6}He{_{gs}}$.
One therefore concludes [3] that the ground state of 
${^8}He$ contains subsystem ${^6}He$ in the excited state $2^+$
with a large weight.

Hence what both these two experiments  are showing is that 
${^8}He$ has a two-tier structure with its ground state having
${^6}He$ in an the excited state ( $2^+$ at 1.8 MeV ).
Therefore the structure of 
${^8}He$ is very different from that of ${^6}He$ which is two loosely 
bound neutrons outside an inert $\alpha$.
But here too though ${^6}He$
is in an excited state still it is commonly believed that the 
$\alpha$ inside it remains inert [1,3].

However this view ignores the fact that in the excited state $2^+$ the 
structure of ${^6}He$ may not remain pure $\alpha$ + 2n but may go over to 
another configuration - that of ${^3}H$ + ${^3}H$ ( ie t+t ). In the past 
there have been several attempts to understand the structure of 
${^6}He$ as consisting both of  $\alpha$ + 2n
and of t+t cluster structures ( see [7] for references about this ). So 
one may justifiably ask if it 
could be that it is the t+t structure of ${^6}He$ that is
manifesting itself 
in this excited state relevant to ${^8}He$. And this is exactly what my 
model predicts [4]. To understand as to how this is possible, first a few 
words about the model.

Recently using QCD and quark model based ideas, the author has pointed
out common threads in such diverse nuclear phenomenon as [4]: 
1. The halo phenomena is neutron rich nuclei,
2. The formation and persistence of clusters in nuclei and
3. The nuclear molecule effects.
4. The hole at the centre of the density distribution in ${^3}H, {^3}He$ 
and ${^4}He$ [5,6].

It was pointed out by the author
that all these require an understanding of two or more 
nucleons strongly overlapping over a small local region of size
$\le 1 fm$ in some specific nuclei. 
This necessarily requires considerations of multi-quark 
configurations like 6-, 9- and 12-quarks. It was shown by the author
[4,5,6] that the nucleons do not like to go into these configurations and 
thereby the above mentioned nuclear effects are given a consistent 
explanation. The reader may refer to [4,5,6] for further details.

Let us here concentrate upon a very significant empirical prediction
of this new model.
That is the effect of triton clustering for neutron rich nuclei. 
It was shown that as more and 
more neutrons are added to light nuclei there is a marked effect that an 
$\alpha$ plus two neutrons prefer to go to a configuration of two tritons.
There is a continuous competition between an $\alpha$ cluster plus two 
neutrons remaining as such versus these changing into two clusters of 
tritons. For a single such configuration in 
${^6}He$ this breakup does not happen and to a 
very good approximation ${^6}He$ remains an $\alpha$ plus two neutrons 
for the ground state [1]. Also
consequences like formation of nuclear molecular structures [4] and 
also providing us with 
a better insight into the structure of other A=6 nuclei like
${^6}Li$ and ${^6}Be$ [7]. The effect of triton
( as well as $\alpha$ ) clustering also 
helps us to explain the phenomenoa of neutron halos [4] and
provides a complete and consistent description of all nuclei known to have 
neutron halo structure [4,8]. It also predicts as to where to look for
further new neutron halo nuclei,

People have always thought that for stable N=Z nuclei with 4n the
alphas give a good cluster structure because of the high
value of binding energy of this particular nucleus.
But the author pointed out [4] that other properties which are significant 
in this respect are the extremely high surface density of $\alpha$
and also that it has a hole at the centre. What the other A=3 nuclei,
${^3}He$ and ${^3}H$ have in common with $\alpha$ are having holes in the 
centre and having very high densities on the surface - much higher than 
any other nuclei [4,5,6]. 
It is these two properties which  all these three light nuclei
( A=4 and A=3 ) have in commom which makes them ideal candidates around 
which cluster structures may develop [4].
This is the reason why the tritons are showing up so prominently in 
forming clusters in neutron rich nuclei. Also for the same reason 
${^3}He$ would play a similar role in proton rich nuclei [4].

This tendency for neutron rich nuclei to form tritons was used 
by the author [4] to predict that 
${^9}Li$ can be treated as made up of 3t and therby forming a 
more compact nucleus. Thus when one goes to 
${^{11}}Li$ one has two loosely bound neutrons forming a loose halo 
around a compact core of ${^9}Li$. However for ${^6}He$ the core 
$\alpha$ remains inert and it is the two valence neutrons which form
the halo. But when one adds two more neutrons around ${^6}He$ to form
${^8}He$ then for the ground state of this nucleus 
there is a distinct change in the substructure of 
${^6}He$ as discussed above. As per our model discussed here, 
The excited $2^+$ state of it can be imagined to have gone into the 
predominant t+t configuration.  This forms a very 
compact structure as its two constituent, the tritons themselves are so 
compact. Therefore it would have high surface densities and thereby 
preventing the
two valence neutrons around it ( for ${^8 }He$ ) from going through and 
thus forming a halo around it. Hence as per our model  
${^8}He$ is also a two neutron halo nucleus with a compact core of
two tritons.

In trying to understand the surprisingly large branching ratio for 
beta-delayed triton emission in ${^8}He$, the authors of ref. [9] suggeted 
the existence of a new $1^+$ state at 9.3 MeV in ${^8}Li$. They did not 
know the origin of this state though. Quite clearly in our model 
here this is the state built up on the $1^+$ state at 5.65 MeV in ${^6}Li$
( see [7] for deatils ). For our purpose here just note
that this excited $1^+$ state in ${^8}Li$ has 
the cluster structure ($\alpha$-t-n) [7]. Therefore 
tritons would be easily emitted in this branch of beta decay in ${^8}He$
whose structure 
as shown here is t+t+2n. The beta transition would connect these two 
structures in a naturally simple manner.

In summary, two recent experimental results are used here to give 
a new understanding
of the structure of the nucleus ${^8}He$. Author's recent model, 
which has been  
successfully applied to provide a comprehensive perspective of several 
diverse nuclear phenomena, again with the
insights obtained from these new 
experiments provides a new description of ${^8}He$.
It is predicted to have a structure of t+t+2n. 
Here the two tritons form a compact core ( as an excited state of 
${^6}He$ ) around which the two valence neutrons form a loose halo. 

\newpage

\vspace{.8in}

{\bf References} 

\vspace{.2in}

1. I.Tanihata, D. Hirata, T. Kobayashi, S. Shimoura, K. Sugimoto and
H. Toki, Phys. Lett. {\bf B289} (1992) 261

\vspace{.2in}

2. R. Wolski, S. I. Sidorchuk, G. M. Ter-Akopian, A. S. Fomichev,
A. M. Rodin, S. V. Stepantsov, W. Mittig, P. Roussel-Chomoz, 
H. Savajols, N. Alamanos, F. Auger, V. Lapoux, R. Raabe, 
Yu. M. Tchuvil'sky and K. Rusek, 
Nucl. Phys. {\bf A 722} (2003) 55c

\vspace{.2in}

3. A. A. Korsheninnikov, Nucl. Phys. {\bf A 722} (2003) 157c

\vspace{.2in}

4. A. Abbas, Mod. Phys. Lett. {\bf A16} (2001) 755.

\vspace{.2in}

5. A. Abbas, Phys. Lett. {\bf B167} (1986) 150. 

\vspace{.2in}

6. A. Abbas, Prog. Part. Nucl. Phys. {\bf 20} (1988) 181.

\vspace{.2in}

\vspace{.1in}

7. A. Abbas, "Structure of A=6 nuclei: ${^6}He$, ${^6}Li$ and ${^6}Be$",
 
IOP Preprint:IP/BBSR/2003-18(June'03): physics/0306186

\vspace{.2in}

8. A. Abbas, "Triton clustering in neutron rich nuclei and Ikeda-like

diagrams",IOP Preprint:IP/BBSR/2003-25(July'03):physics/0307066 

\vspace{.2in}

9. M.J.G. Borge, L. Johannsen, B. Jonson, T. Nilsson, G. Nyman, 
K. Riisager, O. Tengblad and K. Wilhelmsen Rolander, 
Nucl. Phys. {\bf A560} (1993) 664.

\end{document}